
\input harvmac
\def\today{\ifcase\month\or
   January\or February\or March\or April\or May\or June\or
   July\or August\or September\or October\or November\or December\fi
   \space\number\day, \number\year}
%
\def\title#1#2#3#4{\Title{ESENAT-#1}{#2}\vskip -0.3in
\centerline{{\titlefont#3}}\vskip 0.2in
\centerline{{\titlefont#4}}\vskip 0.3in}
%
\def\js{\centerline{Jae-Suk Park\footnote{$^\dagger$}
{Bitnet: pjesenat@krysucc1}}\bigskip
\centerline{{\it Department of Physics, Yonsei University}}
\centerline{{\it Seoul 120-749, Korea}}
\centerline{{\it and}}
\centerline{{\it ESENAT Research Institute for Theoretical Physics}}
\centerline{{\it 70-30 Changcheon, Seoul 120-180,
Korea}\footnote{$^{\dagger\dagger}$}{Present Adress}}\bigskip}
\def\abs#1{\centerline{{\bf Abstract}}\vskip 0.1in
{#1}
}

%
    \def\b{\beta}              \def\d{\delta}
\def\D{\Delta}            \def\F{\Phi}
\def\g{\gamma}               \def\l{\lambda}
   \def\m{\mu}                 
             
\def\P{\Psi}

%
\def\CUG{{\cal U}/{\cal G}}
\def\CA{{\cal A}}   \def\CC{{\cal C}}
 \def\CF{{\cal F}}  \def\CG{{\cal G}}

   \def\CU{{\cal U}}
\def\CW{{\cal W}} \def\CM{{\cal M}}
%
\def\rd{\partial}

\def\darr#1{\raise1.5ex\hbox{$\leftrightarrow$}\mkern-16.5mu #1}
\def\Ha{{1\over2}}

\def\Fr#1#2{{#1\over#2}}
\def\tr{\hbox{Tr}\,}

\def\rF#1{\Fr{\rd}{\rd #1}}

%
\def\cmp#1#2#3{, Comm.\ Math.\ Phys.\ {{\bf #1}} {(#2)} {#3}}
\def\pl#1#2#3{, Phys.\ Lett.\ {{\bf #1}} {(#2)} {#3}}
\def\np#1#2#3{, Nucl.\ Phys.\ {{\bf #1}} {(#2)} {#3}}

\def\prp#1#2#3{, Phys.\ Rep.\ {{\bf #1}} {(#2)} {#3}}

\def\ijmp#1#2#3{, Int.\ J.\ Mod.\ Phys.\ {{\bf #1}} {(#2)} {#3}}

\def\jdg#1#2#3{, J.\ Diff.\ Geo.\ {{\bf #1}} {(#2)} {#3}}
\def\pnas#1#2#3{, Proc.\ Nat.\ Acad.\ Sci.\ USA.\ {{\bf #1}} {(#2)} {#3}}

\def\zp#1#2#3{, Z.\ Phys.\ {{\bf #1}} {(#2)} {#3}}
\def\ap#1#2#3{, Ann.\ Phys.\ {{\bf #1}} {(#2)} {#3}}
\def\ptrsls#1#2#3{, Philos.\ Trans.\  Roy.\ Soc.\ London Ser.\
{{\bf #1}} {(#2)} {#3}}

%
%
\def\dt{\d_{\!{}_{T}}}
\def\dw{\d_{\!{}_{W}}}
\def\db{\d_{\!{}_{B\!R\!S}}}
\def\hCF{\hat{\CF}}
\def\hCA{\hat{\CA}}
\def\tW{\tilde{W}}
\def\tCW{\tilde{\CW}}
\lref\W{E.\ Witten\cmp{117}{1988}{353}\semi
E.\ Witten, Introduction to cohomological field theories,
{\it in\/} Proc.\ Trieste Conference on
Topological methods in quantum field theories (ICTP, Trieste, June
1990), ed.\ W.\ Nahm et.\ al., (World Scientific, Singapore, 1991)
}
\lref\D{S.K.\ Donaldson\jdg{26}{1987}{397}; Topology {\bf 29} (1990) 257
}
\lref\AS{M.F.\ Atiyah and I.M.\ Singer\pnas{81}{1984}{2597}
}
\lref\AHS{M.F.\ Atiyah, N.J.\ Hitchin and
I.M.\ Singer\ptrsls{A308}{1982}{524}
}
\lref\Stora{R.\ Stora, Algebraic structure and topological origin
of anomalies, {\it in\/} Progress in gauge field theory,
ed.\ G.\ 't Hooft et al.\ (Plenum, N.Y., 1984)
}
\lref\H{J.H.\ Horne\np{B318}{1989}{22}
}
\lref\K{H.\ Kanno\zp{C43}{1989}{477}
}
\lref\Park{J.S.\ Park, revised version of ESENAT-/92/04 (hep-th/9204058
to be replaced by August 1992)
}
\lref\BRS{C.\ Becchi, A.\ Rouet and R.\ Stora\cmp{42}{1975}{127}\semi
I.V.\ Tyutin, Lebedev Inst.\ preprint, (1975)
}
\lref\Zumino{B.\ Zumino, Chiral anomalies and differential geometry,
{\it in\/} Relativity, group and topology II, eds.\ B.S.\ deWitt et.\
al., Les Houches 1983, (North-Holland, Amsterdam, 1984)
}
\lref\WZ{ J.\ Wess and B.\ Zumino\pl{B37}{1971}{95}
}
\lref\BSLPBMS{
L.\ Baulieu, I.M.\ Singer\np{(Proc.\ Suppl.) B5}{1988}{12}\semi
J.M.F. Labastida, M.\ Pernici\pl{B212}{1988}{56}\semi
R.\ Brooks, D.\ Montano, J.\ Sonnenschein\pl{B214}{1988}{91}
}
\lref\WMN{
S.\ Wu\pl{B264}{1991}{339}\semi
J.M.\ Maillet and A.J.\ Niemi\pl{B223}{1989}{195}
}
\lref\BBRT{D.\ Birminham, M.\ Blau, M.\ Rakowski and G.\
Thomson\prp{209}{1991}{129}}
\lref\M{R.\ Myers\ijmp{A5}{1990}{1369}
}
\lref\MZ{
J.\ Manes, and B.\ Zumino, Non-triviality of gauge anomalies,
{\it in\/} Supersymmetry and its applications, Proceedings of Nuffield
Workshop on Supersymmetry and its applications (Cambridge, June 1985),
eds.\ G.W.\ Gibbons et.\ al., (Cambridge Univ.\ Press, 1985)
}
\lref\Gri{V.\ Gribov\np{B139}{1978}{1}
}
\lref\Singer{I.M.\ Singer\cmp{60}{1978}{7}
}
\lref\AG{
L.\ Alverez-Gaum\'{e} and P.\ Ginsparg\np{B243}{1984}{449}
}
\lref\Ano{
B.\ Zumino, Y.-S.\ Wu and A.\ Zee\np{B239}{1984}{477}\semi
W.A.\ Bardeen and B.\ Zumino\np{B244}{1984}{421} \semi
B.\ Zumino\np{B253}{1985}{477} \semi
J.\ Manes, R.\ Stora and B.\ Zumino\cmp{102}{1985}{157}\semi
L.\ Alverez-Gaum\'{e} and P.\ Ginsparg\ap{161}{1985}{423}
}
\title{/92/07}{Universal Bundle, Generalized Russian Formula}{and
Non-Abelian Anomaly}{in Topological Yang-Mills Theory}
\vskip 0.3in
\js
\abs{We re-examine the geometry and algebraic structure of BRST's of
Topological Yang-Mills theory based on the universal bundle
formalism of Atiyah and Singer. This enables  us to find  a natural
generalization of the {\it Russian formula and  descent equations\/},
which can be used as algebraic method to find the non-Abelian anomalies
counterparts in Topological Yang-Mills theory. We suggest that the
presence of the non-Abelian anomaly obstructs the proper definition of
Donaldson's invariants.
}

\Date{April 24, 1992; Revised by July 7, 1992}

\newsec{Introduction}
The topological origin of the non-Abelian anomaly have been
investigated by Atiyah and Singer in terms of universal bundle
formalism. Let $M$ denote a $2n-2$ dimensional oriented compact
Riemann manifold.  Consider a principal $G$-bundle $P$ over base
space $M$. Let $\CU$ denote the affine space of all connections
on $P$ and $\CG$ denote bundle automorphism, which acts as the
gauge symmetry group.  Now consider a principal $\CG$-bundle
over base space $(P\times
\CU/\CG)$ where $G$ acts freely; $\left(P\times\CU,
\CG, (P\times\CU)/\CG\right)$.  The base space of the
above bundle itself can be regarded as a principal $G$-bundle
over $M\times\CU/\CG$, which is called the {\it universal
bundle}\AS
\eqn\ub{((P\times \CU)/\CG,G,M\times \CU/\CG).
}
Let $\hCF$ denote curvature form over $M\times\CU/\CG$, then, one
can define a characteristic class given by the rank $n$
invariant polynomial $\tr(\hCF^n)$, which yields a two-form on
$\CU/\CG$ after integrating over $M$. The resulting two-form is
related to the $2n-2$ dimensional non-Abelian anomaly\Stora.  A
systematic algebraic method for obtaining $2n-2$ dimensional
non-Abelian anomaly from $2n$ dimensional Abelian anomaly, which
is given by the rank $n$ invariant polynomial $\tr(F^n)$ on $M$,
was developed using the {\it Russian formula\/} and {\it descent
equation\/}\Stora\Zumino.
And, of course, both methods are closely related.

Recently, the universal bundle formalism has been used for an
another important application to Donaldson-Witten theory of the
smooth four dimensional invariants\D\W.  Let $M$ be a four
dimensional compact oriented Riemann manifold. Now one restricts
the orbit space $\CU/\CG$ to the moduli space $\CM$ of anti-self
dual connections (the universal instanton bundle).  One can
define the invariant second rank polynomial $\tr(\hCF^2)$ which is
an element of the cohomology class $H^4(M\times \CM,R)$. If we
restrict the base space to $Y\times\CM$, where $Y$ is an
$r$-dimensional submanifold of $M$, and by integrating
$\tr(\hCF^2)$ along the fibers of the projection
$Y\times\CM\to\CM$, we can get an element in $H^{4-r}(\CM,R)$
which depends only on the homology class of $Y$, i.e.
$H_r(M,R)$\W. Witten has used the elements $\tW^{4-r_i}\in
H^{4-r_i}(\CM,R)$ as the basic observables of his Topological
Yang-Mills Theory (TYMT in short), which is designed such that
the physical configurations of theory are precisely the
instanton moduli space, and interpreted Donaldson's invariant as
the expectation value of observables
\eqn\eev{
\left<\tW^{4-r_1}\cdots \tW^{4-r_k}\right>,
}
for
\eqn\ess{
\sum^k_{i=1}(4-r_i)=d(\CM),
}
where $d(\CM)$ denote the dimension of moduli space.  In
particular, Witten has introduced fermionic variables
$(\P,\F,\chi)$, whose zero modes precisely correspond to the
cohomology classes of instanton complex\AHS, with $U$-charges
(the ghost numbers of Witten's BRST-like operator $\dw$)
$(1,-1,-1)$ such that
\eqn\edm{
d(\CM)_f = n_{{}_\P}-n_{{}_\F} -n_{{}_\chi},
}
where $d(\CM)_f$ denote the {\it formal\/}
dimension\foot{The formal dimension becomes actual dimension
of moduli space when there are no $\F$ and $\chi$ zero-modes.
The $\F$ and $\chi$ zero-modes are related to the singularities
of moduli space, and Donaldson's invariants are not well-defined
in this case. In particular $\F$ zero-modes arise when there are
reducible connections. We shall discuss this issue later.}
of moduli space and $n_{{}_X}$ denotes the number of  zero-modes
of fermionic fields.
Witten's action has the total ghost number zero and an
observable $\tW^{4-r_i}$ has the ghost number $(4-r_i)$.  Thus the
non-zero dimension of moduli space implies the ghost number
anomaly (an Abelian anomaly) and suitable set of observables
should be inserted according to the superselection rule\ess\ to
compensate the ghost number anomaly\W\WMN.
That is, Donaldson's invariants are manifestations of the Abelian anomaly.

Now it is natural to ask what is the non-Abelian anomaly
counterpart in Donaldson-Witten theory. In this letter we
suggest that the universal bundle formalism of Atiyah and Singer
can provide an answer to the above question. In particular, we propose an
{\it extended Russian formula and descent equation\/} which can be
used to find the
candidates for non-Abelian anomaly counterpart in TYMT. We also
find that the universal bundle formalism can provide an unified
picture of the various BRST algebras; the conventional BRS
algebra\BRS, Witten's $\dw$ algebra, the $\dt$ algebra proposed
by the authors of ref.\BSLPBMS\ and Horne's BRS algebra\H.
Similar formalism had already proposed by Kanno\K\ and
reformulated by Birmingham et.\ al.\BBRT. However, these authors
had failed to uncover the complete structures the various BRST algebras.
And, as the results, the extended Russian formula and descent equation
were absent in their formalism.

\newsec{Universal Bundle and BRST Algebras}

In this section we discuss the origin of the various BRST algebras -
the conventional BRS ($\db$) algebra\BRS\H, Witten's BRST ($\dw$) algebra\W\
and
$\dt$ algebra\BSLPBMS, of TYMT based on the universal bundle formalism\AS.
The presentations of this section are  based on the ref.\Stora,
and we generally follow the conventions of ref.\Zumino.

It is convenient to start from a pull backed bundle $Q$ over
$M\times \CU$ from the universal bundle \ub. One can locally
parametrize $\CU$ by
\eqn\ca{
\CA = g^{-1}A\,g + g^{-1}dg,
}
where $A$ denotes a fixed connection one-form tangent to $M$,
and $g$ is an element of gauge group\foot{We also naturally extend the
action of $g$ to the Lie algebra valued-forms tangent to the orbit space
$\CUG$.} $\CG$ which depends  on
both the space-time coordinates $x^\m$ and some group parameter
$\l^i$.  An arbitrary variation on $\CA$ is
\eqn\deca{
\d \CA = g^{-1} \d A\, g - d_{\CA}(g^{-1}\d g),
}
where we will interpret the operator $\d$ as the exterior
derivative tangent to $\CU$.  If we decompose the operator $\d$
as
\eqn\dec{
\d = \dt + \db,
}
such that $\db$ is the variation (exterior derivation) along
gauge group $\CG$
\eqn\edb{
\db = d\l^i\rF{\l^i},
}
and $\dt$ is the exterior derivative over the orbit space $\CUG$
\eqn\enil{
\d^2 =\dt^2 =\db^2 = \dt\db +\db\dt = 0.
}
Then \deca\ becomes
\eqn\decad{
\d\CA = g^{-1}\dt A\,g -d_\CA(g^{-1}\dt g) - d_\CA
(g^{-1}\db g).
}
such that
\eqn\decada{\eqalign{
\dt\CA &= g^{-1}\dt A\, g - d_{\CA}( g^{-1}\dt g),\cr
\db\CA &= - d_\CA (g^{-1}\db g),\cr}
}
where we have used $\db A=0$.

Introducing the connection one-form on $\CU$
\eqn\cc{-G_{\CA}d^{*}_{\CA}\d\CA\equiv \CC,
}
where
$$
G_{\CA} =\left( d^{*}_{\CA} d_{\CA}\right)^{-1}.
$$
The connection one-form $\CC$ can be also decomposed as
\eqn\cd{\CC = -G_{\CA}d^{*}_{\CA}\d\CA
= - G_{\CA}d^{*}_{\CA}\dt\CA - G_{\CA}d^{*}_{\CA}\db\CA,
}
Then one can define the Faddev-Popov ghost $v$ as
\eqn\cg{
- G_{\CA}d^{*}_{\CA}\db\CA = g^{-1} \db g \equiv v,
}
which is the connection one-form along $\CG$, and the BRS algebra
naturally follows
\eqn\brs{\eqalign{
\db\CA &=- dv - \CA v -v\CA \equiv - d_{\CA}v,\cr
\db v &= - v^2.\cr}
}
The total connection one-form over $M\times \CU$ is
\eqn\tc{
\CA + \CC =\CA - G_{\CA} d_{\CA}^{*} \d\CA,
}
and total curvature over $M\times\CU$ is
\eqn\tf{\eqalign{\hCF
= &(d + \d)(\CA - G_{\CA}d_{\CA}^{*}\d\CA)
+\left(\CA-G_{\CA}d_{\CA}^{*}\d\CA\right)^2\cr
= &\CF+ \left(1-d_\CA G_\CA d^*_\CA\right)\d\CA
- \d\left(G_\CA d^*_\CA\d\CA\right)
+\left(G_\CA d^*_\CA\d\CA\right)^2,\cr }
}
which can be written in components
\eqn\tfc{\eqalign{
&\hCF^{2,0}\equiv\CF=d\CA+\CA^2,\cr
&\hCF^{1,1}=\left(1-d_\CA G_\CA d^*_\CA\right)\d\CA,\cr
&\hCF^{0,2}=-\d\left(G_\CA d^*_\CA\d\CA\right)
+\left(G_\CA d^*_\CA\d\CA\right)^2.\cr }
}
Using the decomposition \decada\ and \decada\cd\cg\brs\ we can get
\eqn\foo{\eqalign{\hCF^{1,1}
& = \left(1-d_\CA G_\CA d^*_\CA\right)(\dt\CA+\db\CA),\cr
& = \left(1-d_\CA G_\CA d^*_\CA\right)\dt\CA,\cr}
}
and
\eqn\fzt{\eqalign{ \hCF^{0,2}
=&-\dt\left(G_\CA d^*_\CA\dt\CA\right)
+\left(G_\CA d^*_\CA\dt\CA\right)^2 +\db v + v^2\cr
&+\dt v -\db(G_\CA d^*_\CA\dt\CA) -\{G_\CA d^*_\CA\dt\CA,v\}\cr
=&-\dt\left(G_\CA d^*_\CA\dt\CA\right)
+\left(G_\CA d^*_\CA\dt\CA\right)^2.\cr}
}
Thus, $\hCF$ is also given by
\eqn\rus{\hCF
= \CF + \left(1-d_\CA G_\CA d^*_\CA\right)\dt A
- \dt\left(G_\CA d^*_\CA\dt\CA\right)
+\left(G_\CA d^*_\CA\dt\CA\right)^2,
}
which means $\hCF$ is the total curvature over $M\times\CU/\CG$.
We can call the above relation an {\it extended Russian formula\/}.
That is, if one restricts the variation of $\CA$ in \deca\ to the gauge
group direction\Stora, the above two equation \tf\rus\ lead to
the well-known {\it Russian formula\/}.

Note that $\left(1-d_\CA G_\CA d^*_\CA\right)$ is the horizontal
projection.  Then
\eqn\hpo{
\hCF^{1,1} = \left(1-d_\CA G_\CA d^*_\CA\right)\dt\CA \equiv \d^H\!\CA,
}
where $\d^H$ denotes the operator of horizontal tangent vector
at $\CU/\CG$. Furthermore, direct calculation  shows that
\eqn\hpop{
\d^H(\d^H\!\CA) = - d_\CA\hCF^{0,2},\qquad \d^H\,\hCF^{0,2} = 0.
}
Being the horizontal variation, $\d^H\!\CA$ should satisfy
\eqn\hc{d^*_\CA(\d^H\!\CA) = 0.}
Applying $\d^H$ to the above condition,  we can get
\eqn\hcc{\eqalign{
\d^H\!\left(d^*_\CA\d^H\!\CA\right)
&= [\d^H\!*\!\CA,\d^H\!\CA] - d^*_\CA(\d^H(\d^H\!\CA))  \cr
&=[\d^H\!*\!\CA,\d^H\!\CA] + d^*_\CA d_\CA\F \cr
&= 0,\cr}
}
which can be read as
\eqn\Fzt{\F=\hCF^{0,2} = -G_\CA[\d^H\!*\!\CA,\d^H\!\CA].}
Thus we have obtained Atiyah-Singer's results\AS.

Note that if we denote $\hCF^{1,1}\equiv\P$, $\hCF^{0,2}=\F$
such that
\eqn\jack{
\hCF = \CF + \Psi + \Phi,
}
and $\d^H\equiv\dw$, we can get  Witten's BRST algebra
\eqn\edw{
\dw\CA = \P,\qquad \dw\P = - d_\CA\F,\qquad\dw\F=0.
}
Let
\eqn\eC{
C \equiv -G_\CA d^*_\CA\dt\CA,
}
such that \cd\ becomes
\eqn\eCC{
\CC = C + v.}
Then \foo\fzt\ lead to the $\dt$ algebra\BSLPBMS
\eqn\edt{\eqalign{
&\dt\CA = \P -d_\CA C,\cr
&\dt C = \F -C^2,\cr
&\dt\P = -[C,\P] - d_\CA\F,\cr
&\dt\F = -[C,\F],\cr }
}
where the last two relation follow from $\dt^2 = 0$.
Using
\eqn\dtp{\dt\CA = \Psi - d_\CA C,
}
one can find that
\eqn\dtpc{
-G_\CA d^*_\CA\dt\CA =-G_\CA d^*_\CA\Psi + C.
}
Note that the following condition is crucial
to the self-consistency of $\dt$ algebra
\eqn\epsit{
d^*_\CA \Psi = 0,}
which is identical to \hc.
We can also find $\db$ algebra from eq.\brs\tfc\foo\fzt
\eqn\edbal{\eqalign{
&\db\CA =  - d_\CA v, \cr
&\db v =  -v^2, \cr
&\db\CF = -[v,\CF], \cr
&\db\Psi =  - [v,\Psi] , \cr
&\db \Phi =  -[v,\Phi],\cr}
}
with
\eqn\eaal{\db C = -\dt v - \{C,v\},
}
which is identical to the additional BRS algebra of ref.\H.
We shall see that this additional BRS structure to $\dt$ and
$\dw$ is crucial for non-Abelian anomalies in TYMT.
Note that we have a natural bi-grading structure of ghost
numbers such that the ghost numbers of the fields
$(\CA,C,\Psi,\Phi,v)$ are $(0,1,1,2,0)$ for $\dt$ algebra and
$(0,0,0,0,1)$ for $\db$-algebra.

Let $\tr(\hCF^n)$ denote an  invariant polynomial
of degree $n$, such that
\eqn\esi{
\db \tr(\hCF^n) =0, \qquad (d +\dt)\tr(\hCF^n) =0,
}
where the first relation follows from \edbal\ and the second one
from the Bianchi identity
\eqn\eBian{
(d + \dt)\hCF + [\CA + C,\hCF]=0.
}
We can expanding the invariant polynomials $\tr(\hCF^n)$ in terms of
the $\dt$ ghost number
\eqn\epvt{\tr(\hCF^n)
=\tCW_{2n}{}^{0,0}+\tCW_{2n-1}{}^{0,1}+\cdots+\tCW_{0}{}^{0,2n},
}
where the superscripts indicate $\db$ and $\dt$ ghost numbers,
respectively, and the subscript indicate the space-time form
degree.
It is well-known that the integration of each terms
$\tCW_{2n-k}{}^{0,k}$
in \epvt\ over  $2n-k$ cycle $\g_{{}_{2n-k}}$ are Witten's observables
for Donaldson's polynomial invariant\W\K;
\eqn\ewo{
\int_{\g_{{}_{2n-k}}}\tCW_{2n-k}{}^{0,k}\equiv \tW^{0,k}.
}
Combining \esi\epvt\ we get
\eqn\esdeca{\eqalign{
d\tCW_{2n}{}^{0,0} &= 0, \cr
d\tCW_{2n-1}{}^{0,1} + \dt\tCW_{2n}{}^{0,0} &= 0, \cr
d\tCW_{2n-2}{}^{0,2} + \dt\tCW_{2n-1}{}^{0,1} &= 0, \cr
\vdots& \cr
d\tCW_{0}{}^{0,2n} + \dt\tCW_{1}{}^{0,2n-1}&= 0, \cr
\dt\tCW_{0}{}^{0,2n} &= 0,\cr}
}
which is identical to the {\it topological
descent equation\/} of ref.\W\K.
Integrating $i$th ($i=2,\ldots,2n$) relation of \esdeca\
over a $2n+2-i$ cycle, we can see that Witten's observables
$\tW_{2n-k}{}^{0,k}$, ($k=0,\ldots,2n$)
are $\dt$ closed;
\eqn\ecdi{
\dt\int_{\g_{{}_{2n-k}}}\tCW_{2n-k}{}^{0,k}
=\dt \tW^{0,k} =0.
}

\newsec{Extended Russian Formula and Descent Equation}

Now  consider the extended Russian
formula of eq.\tf\rus, which can be written as
\eqn\eRusa{\eqalign{
\hCF
= &(d+\dt+\db)(\CA+C+v)+\left(\CA+C+v\right)^2, \cr
= &(d + \dt)(\CA + C) +\left(\CA+C\right)^2,\cr}
}
which reduces to the familiar Russian formula\Stora\Zumino\Ano
\eqn\erus{\CF = (d + \db)(\CA + v) +\left(\CA +v\right)^2
= d\CA + \CA^2,}
for $C=0$, i.e. $(\P=\F=0)$).
Note that $\hCF$ satisfies Bianchi identities
\eqn\eBianc{\eqalign{0
&= (d + \dt)\hCF + [\CA + C,\hCF]\cr
&= (d + \dt +\db)\hCF + [\CA + C + v, \hCF] = 0.\cr}
}
Thus we can see that the invariant polynomial
of degree $n$, $\tr(\hCF^n)$  satisfies
\eqn\esip{
(d + \dt)\tr(\hCF^n) = (d + \dt + \db)\tr(\hCF^n)=0.
}
Then, by the Poincar\'{e} lemma, \esip\ implies
\eqn\epl{\eqalign{\tr(\hCF^n)
&= (d+\dt)\CW_{2n-1}(\CA + C ,\hCF)\cr
&= (d+\dt+\db)\CW_{2n-1}(\CA + C + v,\hCF),\cr}
}
where $\CW_{2n-1}$ denotes the generalized Chern-Simons form.
Expanding $\CW_{2n-1}(\CA + C + v,\hCF)$
with powers of $v$
\eqn\eepv{
\CW_{2n-1}(\CA + C + v,\hCF) = \CW_{2n-1}{}^0(\CA + C,\hCF)
+ \CW_{2n-2}{}^1 + \cdots +\CW_{0}{}^{2n-1},
}
where the superscript indicate the power of $v$ ($\db$ ghost
number) and the subscript indicates the space-time form degree
plus the $\dt$ ghost number.  Then \epl\ reduce   to
\eqn\eedec{\eqalign{
(d+\dt)\CW_{2n-2}{}^1+\db\CW_{2n-1}{}^0&=0,\cr
(d+\dt)\CW_{2n-3}{}^2+\db\CW_{2n-2}{}^1&=0,\cr
\vdots& \cr
(d+\dt)\CW_{0}{}^{2n-1}+\db\CW_{1}{}^{2n-2}&=0,\cr
\db\CW_{0}{}^{2n-1}&= 0.\cr}
}
We shall call the above relations an {\it extended descent
equation}, which is an obvious extension of  the usual
descent equation\Stora\Zumino\Ano\ of Yang-Mills theory
(the original equation can be recovered for $C=0$, i.e. $(\P=\F=0)$).

We can expand the relations of \eedec\ in terms of
$\dt$ ghost number. In particular, consider the second relation of
\eedec
\eqn\enaa{
(d+\dt)\CW_{2n-3}{}^2+\db\CW_{2n-2}{}^1=0,
}
which leads to
\eqn\eenaa{\eqalign{
d\CW_{2n-3}{}^{2,0} +\db\CW_{2n-2}{}^{1,0}&=0,\cr
d\CW_{2n-4}{}^{2,1}+\dt\CW_{2n-3}{}^{2,0}+\db\CW_{2n-3}{}^{1,1}&=0,\cr
d\CW_{2n-5}{}^{2,2}+\dt\CW_{2n-4}{}^{2,1}+\db\CW_{2n-4}{}^{1,2}&=0,\cr
\vdots& \cr
d\CW_{0}{}^{2,2n-3}+\dt\CW_{1}{}^{2,2n-4}+\db\CW_{1}{}^{1,2n-3}&= 0, \cr
\dt\CW_{0}^{2,2n-3}+\db\CW_{0}{}^{1,2n-2}&= 0,\cr}
}
where we have used the expansions in terms of $\dt$ ghost number
$$
\CW_{2n-2}{}^1
= \CW_{2n-2}{}^{1,0} + \CW_{2n-3}{}^{1,1} + \cdots + \CW_{0}{}^{1,2n-2},
$$
$$
\CW_{2n-3}{}^2
= \CW_{2n-3}{}^{2,0} + \CW_{2n-4}{}^{2,1}+\cdots + \CW_{0}{}^{2,2n-3},
$$
such that the superscripts indicate $\db$ and $\dt$ ghost numbers,
respectively, and the subscript indicate the space-time form
degree.

Consider the first relation of \eenaa
$$
d\CW_{2n-3}{}^{2,0}+\db\CW_{2n-2}{}^{1,0} =0,
$$
which leads the well-known Wess-Zumino consistency condition\WZ
\eqn\ewzcc{
\db\int_{\g_{{}_{2n-2}}}\CW_{2n-2}{}^{1,0} =0,
}
that is, $\CW_{2n-2}{}^{1,0}$ gives the non-Abelian anomaly of
local Yang-Mills theory in $2n-2$ dimensions.
Integrating the i-$th$ relation of \eenaa\ over $2n-1-i$
cycle
\eqn\eint{\eqalign{
\db W_{2n-2}{}^{1,0} &=0,\cr
\dt W_{2n-2-\ell}{}^{2,\ell-1} + \db W_{2n-2-\ell}{}^{1,\ell} &=0,
\quad\hbox{for}\quad \ell= 1,\ldots, 2n-2 \cr}
}
where
$$
W_{2n-2-\ell}{}^{1,\ell} = \int_{\g_{{}_{2n-2-\ell}}}\!\!\!\!\!
\CW_{2n-2-\ell}{}^{1,\ell}, \quad
W_{2n-2-\ell}{}^{2,\ell-1} = \int_{\g_{{}_{2n-2-\ell}}}\!\!\!\!\!
\CW_{2n-2-\ell}{}^{2,\ell-1}.
$$
We conjecture that
eq.\eint\ is the extended consistency condition and
$W_{2n-2-\ell}{}^{1,\ell}$ are  the non-Abelian anomaly counterparts in TYMT.

If we expand the first relation of \eedec\ in terms of $\dt$ ghost
number
\eqn\eee{\eqalign{
d\CW_{2n-2}{}^{1,0} +\db\CW_{2n-1}{}^{0,0}&=0,\cr
d\CW_{2n-3}{}^{1,1}+\dt\CW_{2n-2}{}^{1,0}+\db\CW_{2n-2}{}^{0,1}&=0,\cr
\vdots& \cr
d\CW_{0}{}^{1,2n-2}+\dt\CW_{1}{}^{1,2n-3}+\db\CW_{1}{}^{0,2n-2}&= 0, \cr
\dt\CW_{0}^{1,2n-2}+\db\CW_{0}{}^{0,2n-1}&= 0,\cr}
}
which leads
\eqn\ecli{
\dt W_{2n-2-j}{}^{1,j} = -\db W_{2n-2-j}{}^{0,j+1},
}
where $j=0,\dots, 2n-2$.
Thus $W_{2n-2-j}{}^{1,j}$ is $\db$ ($\dt$) closed up to $\dt$
($\db$) exact term.
In addition $W_{2n-2-j}{}^{1,j}$ depends only on the homology class of
$\g$ up to BRST exact term like Witten's observables
$\tW^{0,k}$\W. That is, if $\g_{{}_{2n-2-j}}$ is a boundary,
say $\g_{{}_{2n-2-j}}=\rd\b_{{}_{2n-1-j}}$, then
\eqn\ehomo{\eqalign{W_{2n-2-j}{}^{1,j}
&= \int_{\g_{{}_{2n-2-j}}} \CW_{2n-2-j}{}^{1,j}\cr
&= \int_{\b_{{}_{2n-1-j}}} d\CW_{2n-2-j}{}^{1,j}\cr
&= -\dt\int_{\b_{{}_{2n-1-j}}}\CW_{2n-1-j}{}^{1,j-1}
- \db\int_{\b_{{}_{2n-1-j}}}\CW_{2n-1-j}{}^{0,j}.\cr}
}

Note that the algebraic structures
of both the usual descent equation\Zumino\ and the extended one \eedec\ are
identical if we replace $d$ with $d+\dt$ and $\CA$ with $\CA +C$.
Thus we can follow the same procedure discussed in \Zumino.
By introducing the one-parameter family of connection $1$-forms
over $M\times\CUG$
\eqn\eopo{
\hCA_t = t(\CA + C),
}
and associated field strengths
\eqn\eafs{\eqalign{\hCF_t
&= (d+\dt)\hCA_t + \hCA_t^2 \cr
&= t(d +\dt)(\CA + C) + t^2(\CA +C)^2, \cr}
}
we can obtain the Chern-Weil formula\Zumino
\eqn\ecw{\eqalign{ \tr(\hCF^n)
&= n (d +\dt)\int^1_0 dt\, \tr(\hCA\hCF_t^{n-1}),\cr
&= (d +\dt)\CW_{2n-1}{}^0(\CA + C,\hCF).\cr}
}
Note that $\tr(\hCF^n)$ is also given by
\eqn\ecwa{\eqalign{ \tr(\hCF^n)
&= n (d +\dt+\db)\int^1_0 dt\, \tr((\hCA + v)\hCF_t^{n-1}),\cr
&= (d +\dt+\db)\CW_{2n-1}(\CA + C +v,\hCF).\cr}
}
Now we can explicitly expand $\CW_{2n-1}(\CA + C +v,\hCF)$ in powers
of $v$ as \eepv\ and obtain the first order in $v$
\eqn\efo{
\CW_{2n-2}{}^{1}
= n(n-1)\int^{1}_0 dt(1-t) \hbox{Str} \left(v\, (d+\dt)\left((\CA +
C)\hCF_t^{n-2}\right)\right),
}
where $\hbox{Str}$ denotes the symmetrized trace\Zumino
$$
\hbox{Str}(B_1,B_2,\ldots,B_n) \equiv \sum_{\hbox{Perm.}}\Fr{1}{n!}
\tr (B_{p(1)}\cdots B_{p(n)}).
$$
The next step is to expand \efo\ in terms of $\dt$ ghost number,
and the resulting expressions reduce to $\CW_{2n-2-j}{}^{1,j}$.
To be definite, let us consider the explicit expression for the
non-Abelian anomaly counterpart in TYMT of $4$-dimension, i.e. $n=3$;
\eqn\eexpt{
\CW_{4}{}^1 = \tr\left(v\,(d+\dt)\left((\CA+C)(d+\dt)(\CA + C) +
\Ha(\CA + C)^3\right)\right),
}
which can be expanded in terms of $\dt$ ghost number
%
\eqn\babo{\eqalign{
\CW_{4}{}^{1,0} &= \tr\left[v\,d\!\left(\CA \CF -\Ha\CA^3 \right)\right],\cr
\CW_{3}{}^{1,1} &= \tr\left[v\,
\dt\!\left( \CA\CF -\Ha \CA^3\right)
+ v\,d\!\left( \CA(\Psi -\Ha\{\CA, C\}) + C(\CF -\Ha\CA^2)\right)
\right],\cr
\CW_{2}{}^{1,2} &= \tr\left[v\,
\dt\! \left( \CA(\Psi -\Ha\{\CA, C\}) + C(\CF -\Ha\CA^2)\right)\right.\cr
&\left.\qquad\quad +\, v\,d\!\left( C(\Psi -\Ha\{\CA, C\})
+ \CA(\Phi -\Ha C^2)\right)
\right],\cr
\CW_{1}{}^{1,3} &= \tr\left[v\,
\dt\!\left( C(\Psi -\Ha\{\CA, C\}) + \CA(\Phi -\Ha C^2)\right)
+ v\,d\!\left( C \Phi -\Ha C^3\right)
\right],\cr
\CW_{0}{}^{1,4} &= \tr\left[v\,\dt\!\left(C \Phi -\Ha C^3 \right)\right].\cr}
}
The complete explicit solutions for the extended descent equation
\eedec\ will not be discussed here \Park.

\newsec{Discussion}

We have seen that the various BRST structures of TYMT can be naturally
unified in terms of the universal bundle formalism. In particular,
the BRS ($\db$) structure was crucial in extending {\it Russian formula}
and {\it descent equation}, which enable us to find the non-Abelian
anomaly counterpart in TYMT.

Note that $W_{2n-2-j}{}^{1,j}$ has the $\db$ ghost number $1$, while the
observable for Donaldson's invariant $\tW^{0,k}$ has the $\db$
ghost number $0$. Thus, it is closely related to the
zero-modes of the Faddev-Popov ghost $v$, which exist either for
reducible connections or  the Gribov ambiguity\Gri\M.  This is
due to the obstruction for existence of global cross section of
$\CU\to\CU/\CG$\Singer. The origin of the non-Abelian anomaly
$(W_{2n-2}{}^{1,0})$ of local Yang-Mills theory had already been
interpreted in this way\AG.
Topological Yang-Mills theory has further source of such
obstruction, because it is defined on the space of anti-self
dual connection modulo $\CG$.  Note also that the reducible
connection is a source of singularity in the instanton moduli
space and Donaldson's invariants are not well-defined in this
case\D.  Thus, the generalized non-Abelian anomaly can be
regarded as an obstruction to defining Donaldson's invariants.

\bigskip
I am grateful to J.\ Park and J.M.\ Park for proofreading the manuscript.
\listrefs
\bye